\documentclass[prd,twocolumn,amsmath,showpacs,superscriptaddress,amssymb,floatfix,nofootinbib]{revtex4}

\usepackage{amsfonts,color,xcolor,graphicx,amsfonts,multirow,amsmath}
\usepackage[colorlinks=true,citecolor=blue,linkcolor=blue,urlcolor=blue]{hyperref}

\allowdisplaybreaks
\newcommand{\lag}{\langle}
\newcommand{\rag}{\rangle}
\newcommand{\g}{\gamma_5}
\newcommand{\gmu}{\gamma_\mu}

\newcommand{\an}{a^{\|;n}_{2;a_1}(\mu)}
\newcommand{\xin}{\lag \xi_{2;a_1}^{\|;n}\rag |_\mu }
\newcommand{\xinomu}{\lag \xi_{2;a_1}^{\|;0}\rag|_\mu}

\newcommand{\lab}{\label}

\newcommand{\bgeq}{\begin{eqnarray}}
\newcommand{\edeq}{\end{eqnarray}}

\newcommand{\bgf}{\begin{figure}}
\newcommand{\edf}{\end{figure}}

\newcommand{\bgt}{\begin{table}}
\newcommand{\edt}{\end{table}}

\newcommand{\bgtr}{\begin{tabular}}
\newcommand{\edtr}{\end{tabular}}

\newcommand{\bgfs}{\begin{figure*}}
\newcommand{\edfs}{\end{figure*}}

\newcommand{\bgts}{\begin{table*}}
\newcommand{\edts}{\end{table*}}

\newcommand{\bgc}{\begin{center}}
\newcommand{\edc}{\end{center}}

\newcommand{\ad}{&&\hspace{-0.7cm}}

\begin{document}

\title{$a_1(1260)$-meson longitudinal twist-2 distribution amplitude and the $D\to a_1(1260)\ell^+\nu_\ell$ decay processes}

\author{Dan-Dan Hu}
\author{Hai-Bing Fu}
\email{fuhb@cqu.edu.cn (Corresponding author)}
\author{Tao Zhong}
\email{zhongtao1219@sina.com}
\author{Zai-Hui Wu}
\address{Department of Physics, Guizhou Minzu University, Guiyang 550025, P.R.China}

\author{Xing-Gang Wu}
\email{wuxg@cqu.edu.cn}
\address{Department of Physics, Chongqing Key Laboratory for Strongly Coupled Physics, Chongqing University, Chongqing 401331, P.R. China}

\begin{abstract}
In the paper, we investigate the moments $\langle\xi_{2;a_1}^{\|;n}\rangle$ of the axial-vector $a_1(1260)$-meson distribution amplitude by using the QCD sum rules approach under the background field theory. By considering the vacuum condensates up to dimension-six and the perturbative part up to next-to-leading order QCD corrections, its first five moments at an initial scale $\mu_0=1~{\rm GeV}$ are $\langle\xi_{2;a_1}^{\|;2}\rangle|_{\mu_0} = 0.223 \pm 0.029$, $\langle\xi_{2;a_1}^{\|;4}\rangle|_{\mu_0} = 0.098 \pm 0.008$, $\langle\xi_{2;a_1}^{\|;6}\rangle|_{\mu_0} = 0.056 \pm 0.006$, $\langle\xi_{2;a_1}^{\|;8}\rangle|_{\mu_0} = 0.039 \pm 0.004$ and $\langle\xi_{2;a_1}^{\|;10}\rangle|_{\mu_0} = 0.028 \pm 0.003$, respectively. We then construct a light-cone harmonic oscillator model for $a_1(1260)$-meson longitudinal twist-2 distribution amplitude $\phi_{2;a_1}^{\|}(x,\mu)$, whose model parameters are fitted by using the least squares method. As an application of $\phi_{2;a_1}^{\|}(x,\mu)$, we calculate the transition form factors (TFFs) of $D\to a_1(1260)$ in large and intermediate momentum transfers by using the QCD light-cone sum rules approach. At the largest recoil point ($q^2=0$), we obtain $ A(0) = 0.130_{ - 0.013}^{ + 0.015}$, $V_1(0) = 1.898_{-0.121}^{+0.128}$, $V_2(0) = 0.228_{-0.021}^{ + 0.020}$, and $V_0(0) = 0.217_{ - 0.025}^{ + 0.023}$. By applying the extrapolated TFFs to the semi-leptonic decay $D^{0(+)} \to a_1^{-(0)}(1260)\ell^+\nu_\ell$, we obtain ${\cal B}(D^0\to a_1^-(1260) e^+\nu_e) = (5.261_{-0.639}^{+0.745}) \times   10^{-5}$, ${\cal B}(D^+\to a_1^0(1260) e^+\nu_e) = (6.673_{-0.811}^{+0.947}) \times   10^{-5}$, ${\cal B}(D^0\to a_1^-(1260) \mu^+ \nu_\mu)=(4.732_{-0.590}^{+0.685}) \times   10^{-5}$, ${\cal B}(D^+ \to a_1^0(1260) \mu^+ \nu_\mu)=(6.002_{-0.748}^{+0.796}) \times   10^{-5}$.
\end{abstract}

\date{\today}


\maketitle

\section{Introduction}

The $D$-meson semileptonic decays into axial-vector mesons are the key components for understanding the nonperturbative effects in weak interactions, which have been studied by many theoretical and experimental groups. Based on the constituent quark model, the quantum number of a meson is determined by the quantum numbers of all the constituent quarks. There are two types of axial-vector mesons, $^1P_1$-state with the quantum state $J^{PC} = 1^{+-}$ and $^3P_1$-state with $J^{PC}=1^{++}$. Among the axial-vector mesons, only the isospin triplet heavy state mesons $b_1(1235)$ and $a_1(1260)$ do not have a mixing phenomenon. Their internal structures are relatively clear. At the hadron level, the $a_1(1260)$-meson is a good subject which is considered as the chiral adjoint state of $\rho$-meson. Thus $a_1(1260)$ and $\rho$-meson are defined as the light-quark pair $q \bar q$ with $q=(u,d)$~\cite{Dhar:1983fr, Wingate:1995hy, Wakayama:2019crb}. Since 1986, the properties of $a_1(1260)$-meson have been accurately measured~\cite{Weinstein:1993jp}. Besides, the observation of the charmless hadronic decay processes involving $a_1(1260)$-meson, such as $B^0 \to a_1^\pm(1260)\pi^\mp$, which have been issued by the BABAR and Belle collaborations~\cite{Aubert:2004xg, Aubert:2006dd, Aubert:2006gb, Abe:2007jn}, indicates that $a_1(1260)$ is a $^3P_1$-state. Those measurements help us to investigate the production mechanism of axial-vectors via hadronic decay processes and to probe the structures of axial-vector meson. Thus it is important to give a detailed theoretical study on the semileptonic decay $D\to a_1(1260)\ell^+\nu_\ell$.

The transition form factors (TFFs) of $D\to a_1(1260)$ are key components for investigating the corresponding semileptonic decays. The TFFs for heavy-to-light decay processes have been calculated under various approaches, such as the QCD sum rules (QCDSR)~\cite{Aliev:1999mx}, the covariant light-front quark model (CLFQM)~\cite{Cheng:2003sm, Cheng:2017pcq}, the constituent quark model (CQM)~\cite{Deandrea:1998ww}, the light-cone sum rules (LCSR)~\cite{Wang:2008bw, Yang:2007zt, YanJun:2011rn, Momeni:2016kjz}, the relativistic quark model (RQM)~\cite{Faustov:2019mqr}, the perturbative QCD (PQCD)~\cite{Li:2009tx, Liu:2012jb}, the QCD factorization (QCDF)~\cite{Lu:2021ttf}, the three-point QCD sum rules (3PSR)~\cite{Khosravi:2014hqa}, and etc. Among them, the LCSR approach provides an effective way in determining the non-perturbative parameters of hadronic states. By using the LCSR approach, one can carry out the operator product expansion (OPE) near the light-cone $x^2 \approx 0$, and the nonperturbative hadronic matrix elements can be parameterized as the light-cone distribution amplitudes (LCDAs) with various twists. In the present paper, we shall adopt the LCSR approach to deal with the $D\to a_1(1260)$ TFFs by using a left-handed chiral current, which can highlight the longitudinal leading-twist LCDA contribution.

Within the LCSR approach, the LCDAs not only are basic parameters of hard exclusive processes, but also can reflect the dynamic information of the internal structure of hadrons. One can expand the longitudinal twist-2 LCDA $\phi_{2;a_1}^\|(x,\mu)$ of $a_1(1260)$-meson at an arbitrary scale $\mu$ into a Gegenbauer series, $\phi_{2;a_1}^\|(x,\mu) = 6x\bar x [1+\sum_n \an C_n^{3/2}(\xi)]$, where $\bar x = (1-x)$, $\xi=(2x-1)$ and $\an$ are Gegenbauer moments, whose first nonzero one has been given by Yang at the initial scale $\mu_0 = 1~{\rm GeV}$~\cite{Yang:2007zt}, $a_{2;a_1}^{\|;2}(\mu_0) = -0.02(2)$. This value, along with the higher order of $\an$, can be calculated within the framework of QCDSR under the background field theory (BFTSR)~\cite{Huang:1989gv}. It has been pointed out that the LCDA model based on conformal expansion that is truncated after its first few terms is not suitable for all cases, since the higher-order Gegenbauer terms may have sizable contributions, even if they are generally power suppressed with the increment of $n$ for large momentum transforms. Thus it is important to know more moments for a precise determination of LCDA.

The leading-twist LCDA of a meson can be related to its Bethe-Salpeter wave function. Previous works mainly focused on the wave functions of the pseudoscalar or the vector mesons, cf. Refs.~\cite{Ball:2005ei, Zhong:2018exo, Fu:2014cna, Huang:1994dy, Vega:2009zb}, and there is few research on the axial-vector mesons. In this work, we will construct a light-cone harmonic oscillator (LCHO) model for the $a_1(1260)$-meson longitudinal twist-2 LCDA. The parameters of the model shall be fitted by using the newly calculated moments $\xin$. And we will use the least squares method to do the fitting and to get the optimal solution. For the purpose, we will use BFTSR to calculate the moments $\xin$. Within the framework of BFTSR, the quark and gluon fields are composed by the background fields and their surrounding quantum fluctuations, and the usual vacuum condensates are described by the classical background fields, which provides a clear physical picture for the bound-state internal structures and makes the sum rules calculation more simplified. The BFTSR have been widely used in calculating the LCDAs of the heavy/light mesons~\cite{Zhong:2014jla, Fu:2018vap, Zhong:2016kuv, Huang:2004tp, Zhong:2011rg,Hu:2021zmy}. Here we will adopt this approach to investigate the $a_1(1260)$-meson moments $\xin$ and then provide a more accuracy LCDA $\phi_{2;a_1}^\|(x,\mu)$.

The remaining parts of the paper are organized as follows. In Sec.~\ref{Sec_II}, we present the calculation procedures for the moments of $a_1(1260)$-meson longitudinal twist-2 LCDA, the LCHO model, the TFFs and the branching ratios. Numerical results and discussions are presented in Sec.~\ref{Sec_III}. Section~\ref{Sec_IV} is reserved for a summary.

\section{Calculation Technology}\label{Sec_II}

Firstly, the $a_1(1260)$-meson longitudinal twist-2 LCDA is defined as~\cite{Yang:2007zt}
\begin{align}
&\langle 0 |{\bar q_1}(z){\gamma _\mu }{\gamma _5}{q_2}(-z)|{a_1}(q,\lambda )\rangle
\nonumber\\
&\qquad= i m_{a_1} f^\|_{a_1} \int_0^1 dx e^{i(xz\cdot q - \bar xz\cdot q)} q_\mu \frac{{{e^{*(\lambda )}} \cdot z}}{{q \cdot z}}\phi _{2;{a_1}}^\parallel (x,\mu). \label{Eq:phi2}
\end{align}
Here, two light quarks $q_1 = q_2$, which are $(u, d)$-quark for $a_1(1260)$-meson, respectively. This convention shall be followed throughout the remaining parts of this paper. The $f_{a_1}^\|$ is $a_1(1260)$-meson decay constant, $q$ and ${e^{*(\lambda )}}$ are momentum and polarization vector of $a_1(1260)$-meson. The polarization vector satisfies the relationship $(e^{ * (\lambda )} \cdot z) /({q \cdot z}) \to 1/{m_{a_1}}$~\cite{Ball:2004rg}. By doing the series expansion near $z\to 0$ on both sides of Eq.~\eqref{Eq:phi2}, one will get:
\begin{align}
\langle 0|{\bar q_1}(0)/\!\!\! z\g  {(iz\cdot\tensor D)^{n}}{q_2}(0)|& a_1(q,\lambda )\rangle
\nonumber\\
&=i{(z\cdot q)}^{n+1} f_{a_1}^\| \langle\xi^{n;\|}_{2;a_1}\rangle|_\mu,
\end{align}
where the covariant derivative satisfies the relation $(iz \cdot \tensor D)^n=(iz\cdot \overrightarrow D - iz\cdot \overleftarrow D)^n$. The $n$-th order moments of the $a_1(1260)$-meson DA are defined as
\begin{align}
\langle \xi _{2;{a_1}}^{\parallel ;n}\rangle {|_\mu } = \int_0^1 dx{{{(2x - 1)}^n}} \phi _{2;{a_1}}^\parallel (x,\mu )
\end{align}
One can start from the following correlation function (correlator) to derive the sum rules, i.e.
\begin{eqnarray}
\Pi_{2;a_1}^{(n,0)}(z,q) &=&i \int d^4x e^{iq\cdot x}\langle 0 |T\{ J_n(x),J_0^\dagger(0)\}|0\rangle
\nonumber \\
&=&(z\cdot q)^{n+2} I_{2;a_1}^{(n,0)}(q^2), \label{Eq:Correlator1}
\end{eqnarray}
with $J_n(x)= \bar q_1(x)/\!\!\! z\g (iz\cdot\tensor D)^n q_2(x)$, $J_0^\dagger(0)= \bar q_2(0)/\!\!\! z\g q_1(0)$ and $z^2=0$.\footnote{Here, the current $J_n(x)$ mainly comes from the basic definition of $\phi_{2; ^3P_1}^{\|} (x,\mu)$, which is also in agreement with the current in calculating the Gegenbauer moments of $^3P_1$ axial-vector meson, i.e. $a_{2;^3P_1}^{\|,n}(\mu)$ from Ref.~\cite{Yang:2007zt}. The slightly difference lies in the relationship between two types of moments.} Because of the G-parity, $\phi _{2;a_1}^\|(x,\mu)$ for $^3P_1$-state defined by the nonlocal axial-vector current is symmetric, indicating only even moments are non-zero, i.e. $n=(0,2,4,6,\cdots)$. Based on the idea of BFTSR and the Feynman rules for one hand, one can apply the OPE for the correlator~(\ref{Eq:Correlator1}) in deep Euclidean region $q^2 \ll 0$. Then, the correlator can be expanded into three terms including the quark propagators $S_F^d(0,x)$, $S_F^u(x,0)$ and the vertex operators $(iz\cdot \tensor{D})^{n}$, which have been given in our previous work~\cite{Hu:2021zmy}. In dealing with Lorentz invariant scalar function $\Pi_{2;a_1}^{(n,0)}(z, q^2)$, the vacuum matrix element should be used, which can be found in Ref.~\cite{Zhong:2014jla}. On the other hand, one can insert a complete set of $a_1(1260)$-meson intermediated hadronic states with the same $J^P$ quantum number into the correlator and obtain
\begin{eqnarray}
\text{Im} I_{2;a_1,{\rm Had}}^{(n,0)}(q^2)&=&\pi \delta(q^2-m_{a_1}^2) f_{a_1}^2 \langle \xi^{\|;n}_{2;a_1} \rangle|_\mu \xinomu
\nonumber\\
&+&\frac{3}{4\pi(n+1)(n+3)}\theta (q^2-s_{a_1}),  \label{rm}
\end{eqnarray}
where $s_{a_1}$ is the continuum threshold. The first (second) terms on the r.h.s of Eq.~\eqref{rm} are the $a_1(1260)$-meson ground state (continuum states) contribution. Due to not all higher-order contributions and higher-dimensional operators have been included, the fixed-order prediction of $\xinomu$ is close but not exactly equals to $1$. So we reserve the term $\xinomu$ in the hadronic expression, i.e. Eq.~\eqref{rm}, which is different from other researches in dealing with the axial-meson DA by using the QCDSR approach. Then, one can bridge the invariant function and the OPE side by using the dispersion relation. Furthermore, the Borel transformation are used to suppress the contribution from the continuum states and high dimension condensates. The sum rule expression is
\begin{eqnarray}
\frac{1}{\pi M^2}\int ds e^{-s/M^2}\textrm{Im}I_{a_1,\textrm{had}}^{(n,0)}(s)=\hat {\cal B}_{M^2} I_{2;a_1,{\rm QCD}}^{(n,0)}(q^2),
\label{Eq:dispersion relation}
\end{eqnarray}
where the Borel parameter $M^2$ coming from the Borel transformation with the operator $\hat{\cal B}_{M^2}$. Following the standard SVZ sum rules procedures, we then obtain the expression of the moment of $a_1(1260)$-meson longitudinal twist-2 LCDA, i.e.,
\begin{widetext}
\begin{eqnarray}
\ad~~\frac{f_{a_1}^{\|2} \langle \xi^{\|;n}_{2;a_1}\rangle|_\mu \langle \xi^{\|;0}_{2;a_1}\rangle|_\mu}{M^2 e^{m_{a_1}^2/M^2}} =\frac{3}{4\pi^2 (n+1)(n+3) }\bigg(1 + \frac{\alpha_s}{\pi} A'_n\bigg)(1 - e^{-{s_{a_1}}/M^2}) +\frac{(m_u + m_d)\langle \bar qq\rangle }{M^4}+\frac{\langle \alpha_s G^2\rangle }{12\pi M^4}\frac{1 + n\theta (n - 2)}{n + 1}
\nonumber \\
\ad~~\qquad -\frac{8n + 1}{18}~\frac{{(m_u + m_d)}\langle  g_s \bar q\sigma TGq\rangle }{M^6} +\frac{\langle  g_s \bar qq\rangle^2 }{81{M^6}}4(2n + 1)-\frac{\langle g_s^3f{G^3}\rangle }{48\pi^2 M^6}~n\theta (n - 2)~+\,\frac{\langle g_s^2\bar qq\rangle^2}{M^6}~\frac{2+\kappa^2}{486\pi^2} \bigg\{ 3\,(17n + 35)
\nonumber \\
\ad~~\qquad -2(51n +25)\bigg(-\ln\frac{M^2}{\mu^2}\bigg)+ \theta (n-2)\bigg[ 2n\bigg(-\ln\frac{M^2}{\mu^2}\bigg) - 25(2n + 1)~\tilde \psi (n) +\frac1n (49{n^2} + 100n+ 56)\bigg]\bigg\}.   \lab{xi2}
\end{eqnarray}
\end{widetext}
Here, $\tilde\psi(n)=\psi((n+1)/2)-\psi (n/2)+\ln4$. The next-to-leading (NLO) corrections are $A'_0=0$, $A'_2=5/3$, $A'_4=59/127$, $A'_6=353/135$~\cite{Ball:1996tb}. When taking $n=0$ for the Eq.~\eqref{xi2}, one can get the sum rule of zeroth moment, e.g. $\langle \xi_{2;a_1}^{\|;0}\rangle|_\mu$. To derive more accurate $\xin$, we adopt
\begin{equation}
 \xin = \frac{(\xin\xinomu)\Big|_{\text{From Eq.~\eqref{xi2}}}}{\sqrt{(\langle \xi_{2;a_1}^{\|;0}\rangle|_\mu)^2}},
 \label{Eq:xin}
\end{equation}
where, the numerator of Eq.~\eqref{Eq:xin} are coming from the Eq.~\eqref{xi2}, and denominator are the zeroth moment.

Owning to the fact that the high-order Gegenbauer moment for $a_1(1260)$-meson longitudinal twist-2 DA still have large uncertainties, one can construct a new LCDA model, i.e. the LCHO model based on the Brodsky-Huang-Lepage (BHL) prescription~\cite{Guo:1991eb,Huang:1994dy}. The BHL suggested a connection between the equal-time wave function in the rest frame and the light-cone wave function by equating the off-shell propagator $\epsilon$ in the two frames. For the former $\epsilon = M^2 - \left( \sum_{i=1}^n q_i^0 \right)^2$ with $\sum_{i=1}^n \textbf{q}_i = 0$, for the latter $\epsilon = M^2 - \sum_{i=1}^n \left[ (\textbf{k}_{\perp i}^2 + m_i^2)/x_i \right]$ with $\sum_{i=1}^n \textbf{k}_{\perp i} = 0$ and $\sum_{i=1}^n x_i = 1$. In the two-particle system, one has
\begin{align}
\textbf{q}^2 \longleftrightarrow \frac{\textbf{k}_\perp^2 + m_q^2}{4x(1-x)} - m_q^2,
\end{align}
with $m_1 = m_2 = m_q$. Then the possible connection between the rest frame wave function $\psi_{\rm CM}(\textbf{q}^2)$ and the light-cone wave function $\psi_{\rm LC}(x, \textbf{k}_\perp)$ can be formally represented by
\begin{align}
\psi_{\rm CM}(\textbf{q}^2) \longleftrightarrow \psi_{\rm LC}\left( \frac{\textbf{k}_\perp^2 + m_q^2}{4x(1-x)} - m_q^2 \right).
\label{CMLC}
\end{align}
On the other hand, the wave function of the harmonic oscillator model in the rest frame is
\begin{align}
\psi_{\rm CM}(\textbf{q}^2) = A \exp \left( - \frac{\textbf{q}^2}{2\beta^2} \right),
\label{CMWF}
\end{align}
from an approximate bound state solution in the quark models for mesons. By combining Eqs.~\eqref{CMLC} and \eqref{CMWF}, the LCHO model of $a_1(1260)$-meson wave function satisfies
\begin{align}
\Psi_{2;a_1}^\|(x,{\mathbf k_\bot})\propto \exp \bigg[ -\frac{{\mathbf k_ \bot ^2 + m_q^2}}{8\beta_{2;a_1}^2x\bar x}\bigg].
\label{Eq:Psi}
\end{align}
Then the LCHO model for the $a_1(1260)$-meson wave function gives
\begin{align}
\Psi_{2;a_1}^\|(x,{\mathbf k_\bot})&=  A_{2;a_1}^\| \varphi _{2;a_1}^\| (x)\exp\bigg[ -\frac{{\mathbf k_ \bot ^2 + m_q^2}}{8\beta_{2;a_1}^2x\bar x}\bigg],
\end{align}
where $A_{2;a_1}^\|$ is the normalization constant, $\beta_{2;a_1}$ is a harmonic parameter, and $m_q$ is the mass of the constitute quark $u$ and $d$ in $a_1(1260)$-meson. In addition, the function $\varphi _{2;a_1}^\| (x)$ dominates longitudinal distribution and can be expressed as~\cite{Zhong:2021epq}
\begin{align}
\varphi _{2;a_1}^\| (x) = (x\bar x)^{\alpha _{2;a_1}^\|}[1 + B_{2;a_1}^\| C_2^{3/2}(\xi)].
\end{align}
Since the meson LCDA is related to its wave function $\Psi_{2;a_1}^\|(x,{\mathbf k_\bot})$ via the following relation:
\begin{align}
\phi _{2;a_1}^\| (x,\mu ) = \frac{2\sqrt 6 }{f_{a_1}^\|}\int_{|{\bf k}_\bot{|^2} \le {\mu ^2}} {\frac{{d^2}{\bf k}_\bot}{16\pi^3}}\Psi _{2;a_1}^\|(x,{\mathbf k_\bot }).
\end{align}
Then, the twist-2 LCDA of $a_1(1260)$-meson can be derived by integrating the transverse momentum, which have the following form
\begin{align}
&\hspace{-0.2cm}\phi _{2;a_1}^\| (x,\mu ) = \frac{\sqrt 6 A_{2;a_1}^\| \beta _{2;a_1}^2}{{\pi ^2}{f_{a_1}^\|}}x\bar x\varphi _{2;a_1}^\| (x)
\nonumber \\
&~~~~~ \times \bigg\{\!\exp\!\bigg[\! - \frac{m_q^2}{8\beta _{2;a_1}^2x\bar x}\bigg] - \exp\bigg[ - \frac{m_q^2 + {\mu ^2}}{8\beta_{2;a_1}^2x\bar x}\bigg]\bigg\}.
\label{Eq:phi}
\end{align}
The two parameters $A_{2;a_1}^\|$ and $\beta _{2;a_1}$ are constrained by the following two conditions:
\begin{itemize}
  \item The wave function normalization condition,
      \begin{eqnarray}
       \int_0^1 dx \int \frac{d^2{\mathbf k_\bot}}{16\pi^3} \Psi_{2;a_1}^\|(x,{\bf k}_\bot) = \frac{f_{a_1}^\|}{2\sqrt 6}.
       \label{Eq:nc}
\end{eqnarray}
  \item The probability of $|q\bar q\rangle$ Fock state in a $a_1(1260)$-meson should be less than $1$, e.g. $P_{a_1}<1$,
      \begin{align}
       P_{a_1} &= \int_0^1 dx \int \frac{d^2{\bf k}_\bot}{16\pi^3}|\Psi_{2;a_1}^\|(x,{\bf k}_\bot)|^2
       \nonumber \\
       &= \frac{(A_{2;a_1}^\|)^2 \beta _{2;a_1}^2}{4{\pi ^2}}\int_0^1 dx [\varphi_{2;a_1}^\|(x)]^2x\bar x
       \nonumber \\
       &\times \exp\bigg[- \frac{m_q^2}{4x\bar x\beta _{2;a_1}^2}\bigg].
       \label{Eq:qq}
      \end{align}

\end{itemize}
We shall fit the parameters $\alpha _{2;a_1}^\|$ and $B_{2;a_1}^\|$ by using the least squares method so as to achieve the same moments $\xin$ from the sum rules (\ref{xi2}). The detailed analysis about this point can be found in Refs.\cite{Zhong:2021epq, Zhong:2022ecl}.

Secondly, we adopt the following correlator to derive the LCSRs for the $D \to a_1(1260)$ TFFs,
\begin{align}
\Pi_\mu (p,q) =& i\int d^4x e^{iq\cdot x}\langle a_1(p,\lambda) |T\{j_\mu(x),j_D^\dag(0)\}|0\rangle
\nonumber \\
=&  -\Pi_1 e_\mu^{*(\lambda )} + \Pi_2(e^{*(\lambda)}\cdot q)(2p+q)_\mu
\nonumber \\[1ex]
&+ \Pi_3(e^{*(\lambda)}\cdot q){q_\mu } + i{\Pi _V} \epsilon_\mu^{\alpha\beta\gamma} e_\alpha^{*(\lambda )} q_\beta p_\gamma,
\lab{cf}
\end{align}
where $j_\mu(x)=\bar q_2(x)\gmu (1-\g )c(x)$ and $j_D^\dag (x) = i\bar c(x)(1 - {\g })q_1(x)$. In the time-like $q^2$-region, the long distance quark-gluon interactions are dominant. To deal with the correlator in the time-like region, one can insert a complete set of the $D$-meson states, which have the same $J^P$ quantum numbers to obtain the hadronic expression. After separating the $D$-meson pole term, we obtain
\begin{align}
&\hspace{-0.35cm}\Pi_\mu^{\rm H}(p,q)  = \frac{{\langle a_1|\bar q_2 \gamma_\mu (1-\gamma_5) c |D \rangle \langle D|\bar ci \gamma_5 q_1|0\rangle }}{m_D^2 - (p+q)^2}
\nonumber \\
&\quad +\sum\limits_{\rm H}\frac{\langle a_1|\bar q_2\gamma_\mu (1-\gamma_5) c|D^{\rm H}\rangle \langle D^{\rm H}|\bar ci(1-\gamma_5)q_1|0\rangle }{m_{D^{\rm H}}^{2} - (p+q)^2},
\end{align}
where $\langle D|\bar c i\g q|0\rangle = m_D^2 f_D/m_c$. The $D\to a_1(1260)$ transition matrix elements have the expressions~\cite{Li:2009tx}:
\begin{eqnarray}
&&\hspace{-0.45cm}\langle a_1(p,\lambda)|\bar q_2 \gamma_\mu \gamma_5  c|D(p+q)\rangle \!=\! -\epsilon^{\mu\nu\alpha\beta}\!e_\nu^{*(\lambda )}\!q_\alpha p_\beta
\frac{2iA(q^2)}{m_D- m_{a_1}},
\nonumber\\[1ex]
\\\lab{Eq:DefA}
\nonumber\\
&&\hspace{-0.45cm}\langle a_1(p,\lambda)|\bar q_2 \gamma_\mu c|D(p+q)\rangle ~= - e_\mu ^{*(\lambda )}~(m_D~- m_{a_1})~V_1(q^2)
\nonumber \\
&&\qquad+ (2p+q)_\mu \frac{e^{*(\lambda)} \cdot q}{m_D- m_{a_1}} V_2(q^2)
+ q_\mu (e^{*(\lambda)}\cdot q) \frac{2m_{a_1}}{q^2}
\nonumber \\
&&\qquad\times  [V_3(q^2)- V_0(q^2)],
\nonumber\\
\end{eqnarray}
where $p$ is $a_1(1260)$-meson momentum and $q=p_D-p_{a_1}$ is the momentum transfer, ${e^{*(\lambda )}}$ stands for $a_1(1260)$-meson polarization vector with $\lambda = (\bot,\|)$ being its transverse or longitudinal component, respectively. There are one linear relationships among the TFFs~\cite{Wang:2008bw, Verma:2011yw}:
\begin{align}
\ad V_3(q^2)= \frac{m_D- m_{a_1}}{2m_{a_1}} V_1(q^2) - \frac{m_D+ m_{a_1}}{2m_{a_1}} V_2(q^2),
\end{align}
Following the standard sum rules procedures, one can represent the contributions of the higher resonances and the continuum states by dispersion integrations so as to derive the expressions for the hadronic invariant amplitudes $\Pi_{i}^{\rm H}[q^2,(p+q)^2]$ with $i = (1,2,3,V)$ defined in Eq.~(\ref{cf}). The continuum threshold parameter $s_0$ can be set as the value close to the squared mass of the lowest scalar $D$-meson. Meanwhile, the conventional quark-hadron duality ansatz, $\rho_{i}^\textrm{had} = \rho_{i}^\text{QCD}\theta(s-s_0)$, can be used to calculate the hadron spectrum density $\rho_{i}^\textrm{had}$. On the other hand, in the space-like region, one can calculate the correlator via the QCD theory. In this region, the correlator can be treated by the OPE with the coefficients being pQCD calculable. The $c$-quark propagator which shall be used in the calculation can be found in Ref.~\cite{Hu:2021zmy}. After applying the OPE and using the expressions for the transition matrix elements, one can arrange the resultant expressions by twist-2, 3, 4 LCDAs~\cite{Yang:2007zt, Momeni:2016kjz}. After matching the correlator with the dispersion relation, and applying the conventional Borel transformation to suppress the less known continuum contributions, the resultant TFFs under the LCSR approach are
\begin{widetext}
\begin{eqnarray}
\ad V_1(q^2) = \frac{2m_c^2m_{a_1}f_{a_1}^\|}{m_D^2f_D(m_D - m_{a_1})} \int_0^1\frac{du}{u} e^{(m_D^2-s(u))/{\cal M}^2} \bigg[\Theta(c(u,s_0))\phi_{3;a_1}^\bot (u) - \widetilde\Theta(c(u,s_0))\frac{m_{a_1}^2}{u{\cal M}^2}\Psi_{4;a_1}^\|(u)\bigg] + \frac{2m_c^2m_{a_1}^2}{m_D^2 f_D}
\nonumber \\
\ad \qquad\quad\times \frac{(f_{3;a_1}^V - f_{3;a_1}^A)}{(m_D- m_{a_1})}\int {{\rm{\cal D}}\alpha_i } \int_0^1 dv e^{(m_D^2-s(X))/{\cal M}^2}  \frac1{X^2{\cal M}^2} \Theta(c(X,s_0)) \left[\tilde \Phi_{3;a_1}^\|(\alpha_i)-\Phi_{3;a_1}^\|(\alpha_i)\right],
\label{Eq:V1q2}
\\
\ad V_2(q^2) = \frac{2m_c^2m_{a_1}f_{a_1}^\|(m_D- m_{a_1})}{m_D^2f_D}\int_0^1 \frac{du}{u}e^{(m_D^2-s(u))/{\cal M}^2} ~\bigg[ \frac1{u {\cal M}^2} \widetilde\Theta(c(u,s_0)) \Phi_{2;a_1}^\|(u) + \frac{m_{a_1}^2}{u{\cal M}^4} \widetilde{\widetilde\Theta}(c(u,s_0))
\Psi_{4;a_1}^\|(u) \bigg]
\nonumber \\
\ad \qquad\quad -\frac{m_c^2m_{a_1}^2(f_{3;a_1}^V \!-\! f_{3;a_1}^A)(m_D\!-\! m_{a_1})}{m_D^2 f_D}\int \!{\cal D}\alpha_i \!\int_0^1 \! dv e^{(m_D^2-s(X))/{\cal M}^2} \frac1{X^3 {\cal M}^4} \Theta(c(X,s_0)) \bigg[\tilde \Phi_{3;a_1}^\| (\alpha_i )- \Phi _{3;a_1}^\| (\alpha_i )\bigg],
\label{Eq:V2q2}
\\
\ad  V_0(q^2)=V_3(q^2)+\bigg\{ \frac{q^2m_c^2f_{a_1}^\|}{m_D^2 f_D}\int_0^1 \frac{du}{u} e^{(m_D^2- s(u)) / {\cal M}^2} \bigg[\frac1{u{\cal M}^2} \widetilde\Theta(c(u,s_0)) \Phi_{2;a_1}^\|(u) \!-\! \frac{m_{a_1}^2(2-u)}{u^2 {\cal M}^4}\widetilde{\widetilde\Theta}(c(u,s_0)) \Psi_{4;a_1}^\|(u) \bigg]
\nonumber \\
\ad \qquad\quad -\frac{q^2\,m_c^2\,{\color{blue}m_{a_1}}\,(f_{3;a_1}^V - f_{3;a_1}^A)}{m_D^2{f_D}}\int {\cal D}\alpha_i \int_0^1 dv ~e^{(m_D^2-s(X))/{\cal M}^2}~\frac1{X^3{\cal M}^4} ~\Theta(c(X,s_0))~\bigg[\,\tilde\Phi_{3;a_1}^\| (\alpha_i)\,-\, \Phi_{3;a_1}^\|(\alpha_i )\,\bigg]\bigg\},
\label{Eq:V3q2}
\\
\ad A(q^2)=\frac{m_c^2m_{a_1}f_{a_1}^\|(m_D- m_{a_1})}{2m_D^2 f_D}\int_0^1 du e^{(m_D^2-s(u))/{\cal M}^2} \frac1{u^2{\cal M}^2}\widetilde\Theta(c(u,s_0))\psi_{3;a_1}^\bot (u),
\label{Eq:Aq2}
\end{eqnarray}
\end{widetext}
where the $q^2$-dependence factors $s(u)$ and $s(X)$ are defined as
\begin{align}
s(\zeta)&=&\frac{m_c^2 - (1 - \zeta)({q^2} - \zeta m_{a_1}^2)}{\zeta} \text{ with } \zeta = (u,X),
\end{align}
where $X=\alpha _1 -\alpha _2 + v \alpha _3$, $\alpha_1, \alpha _2$ and $\alpha_3$ are the respective momentum fractions carried by ${\bar q}_1, q_2$ quarks and gluon in the $a_1(1260)$-meson~\cite{Yang:2007zt}. $\Theta(c(u,s_0))$ is the conventional step function, $\widetilde\Theta (c(u,s_0))$ and $\widetilde{\widetilde\Theta}(c(u,s_0))$ are defined as
\begin{align}
&\int_0^1 \frac{du}{u^2 {\cal M}^2} e^{-s(u)/{\cal M}^2}\widetilde\Theta(c(u,s_0))f(u)
\nonumber\\
&\qquad= \int_{u_0}^1\frac{du}{u^2 {\cal M}^2} e^{-s(u)/{\cal M}^2}f(u) + \delta(c(u_0,s_0)),
\label{Theta1}\\
&\int_0^1 \frac{du}{2u^3 {\cal M}^4} e^{-s(u)/{\cal M}^2}\widetilde{\widetilde\Theta}(c(u,s_0))f(u)
\nonumber\\
&\qquad= \int_{u_0}^1 \frac{du}{2u^3 {\cal M}^4} e^{-s(u)/{\cal M}^2}f(u)+\Delta(c(u_0,s_0)), \label{Theta2}
\end{align}
where $c(u,s_0)=u s_0 - m_b^2 + \bar u q^2 - u \bar u m_{a_1}^2$ and
\begin{align}
&\delta(c(u,s_0))= e^{-s_0/{\cal M}^2}\frac{f(u_0)}{m_c^2 + u_0^2 m_{a_1}^2 - q^2}, \\
&\Delta(c(u,s_0))= e^{-s_0/{\cal M}^2}\bigg[\frac{1}{2 u_0 {\cal M}^2}\frac{f(u_0)} {m_c^2 + u_0^2 m_{a_1}^2 - q^2} \nonumber\\
&\left. -\frac{u_0^2}{2(m_c^2 + u_0^2 m_{a_1}^2 - q^2)} \frac{d}{du}\left( \frac{f(u)}{u(m_c^2 + u^2m_{a_1}^2 - q^2)} \right) \right|_{u = {u_0}}\bigg].
\end{align}
Here $u_0$ is the solution of $c(u_0,s_0) = 0$ with $0\leq u_0 \leq 1$. The simplified LCDAs are defined as
\begin{align}
&\Phi_{2;a_1}^\| (u) = \int_0^u dv\phi _{2;a_1}^\| (v),
\\
&\Phi_{3;a_1}^\bot (u) = \int_0^u {dv} \phi_{3;a_1}^\bot (v),
\\
&\Psi_{4;a_1}^\|(u) = \int_0^u {dv} \int_0^v {dw} \psi _{4;a_1}^\|(w).
\end{align}
The coupling constants $f_{3;a_1}^V$ and $f_{3;a_1}^A$ for $a_1(1260)$-meson are defined as the following matrix elements:
\begin{align}
&\langle a_1(q,\lambda)|J_{3,\mu}^{3,A}(0)|0 \rangle = -f_{3;a_1}^A(z\cdot q)^3e^{(\lambda)}_{\bot,\mu}+{\cal O}(z_\mu),\nonumber \\
&\langle a_1(q,\lambda)|J_{3,\mu}^{1,V}(0)|0 \rangle = -if_{3;a_1}^V(z\cdot q)^2e^{(\lambda)}_{\bot,\mu}+{\cal O}(z_\mu),
\end{align}
where the interpolating currents, $J_{3,\mu}^{3,A}(0)=z^{\alpha}z^{\beta}\bar q_2(0)\gamma_{\alpha}\gamma_5[G_{\beta\mu}(0)i(z\cdot \overrightarrow D)- i(z\cdot \overleftarrow D) G_{\beta\mu}(0)]q_1(0)$, $J_{3,\mu}^{1,V}(0)=z^{\alpha}z^{\beta}\bar q_2(0)\gamma_{\alpha}g_s{\tilde G_{\beta \mu }}(0)q_1(0)$, and ${\cal O}(z_\mu)$ involve the twist-4 corrections~\cite{Yang:2007zt}.

Then, the longitudinal and transverse differential decay widths for semileptonic decay $D\to a_1(1260)\ell^+\nu_\ell$ can be expressed as
\begin{align}
&\hspace{-0.45cm}\frac{d\Gamma_{\rm L}(D\to a_1(1260)\ell^+\nu_\ell)}{dq^2}
= \bigg(\frac{q^2 - m_\ell^2}{q^2}\bigg)^2\frac{\sqrt\lambda G_F^2|V_{cd}|^2}{384\pi ^3m_D^3}
\nonumber \\
&\times\frac{1}{q^2}\bigg[3m_\ell^2\,\lambda\,V_0^2(q^2) + \frac{m_\ell^2 + 2q^2}{2m_{a_1}}\Big|(m_D^2 - m_{a_1}^2- q^2)
\nonumber \\
&\times  (m_D- m_{a_1})V_1(q^2)- \frac{\lambda}{m_D- m_{a_1}}{V_2}(q^2)\Big|^2\bigg],
\nonumber\\
\label{Eq:dgamma1}
\\
&\hspace{-0.45cm}\frac{d\Gamma_\pm(D\!\to\! a_1(1260)\ell^+\nu_\ell)}{dq^2}\!  =\! \bigg(\frac{q^2 - m_\ell^2}{q^2}\bigg)^2\frac{\lambda^{3/2}  G_F^2|{V_{cd}}|^2}{384\pi^3m_D^3}
\nonumber \\
&\times(m_\ell^2 + 2q^2)\bigg|\frac{A(q^2)}{m_D- m_{a_1}}\,\mp\,\frac{(m_D- m_{a_1})\,V_1(q^2)}{\sqrt \lambda }\bigg|^2,
\nonumber\\
\label{Eq:dgamma2}
\end{align}
where $G_F$ is the Fermi coupling constant, $|V_{cd}|$ is the CKM matrix element, and $\lambda  = (m_D^4 + m_{a_1}^4 + q^4 - 2m_{a_1}^2m_D^2 -2q^2m_{D}^2- 2q^2m_{a_1}^2)$. The total differential decay width of the semileptonic decay is $d\Gamma_{\rm L}+d\Gamma_{\rm T}$, where $d\Gamma_{\rm L}$ and $d\Gamma_{\rm T}=d\Gamma_++d\Gamma_-$ corresponds to longitudinal and transverse parts, respectively.

\section{Numerical results and discussions}\label{Sec_III}

To do the numerical calculation, the input parameters are taken as follows. The current charm-quark mass, ${\bar m_c}({\bar m_c}) = 1.27(2)~ {\rm GeV}$, and the current quark-masses for the light quarks are ${\bar m_u}(2~{\rm GeV}) = 2.16_{-0.26}^{+0.49}~{\rm MeV}$ and ${\bar m_d}(2~{\rm GeV}) = 4.67_{-0.17}^{+0.48}~{\rm MeV}$. The $D$-meson and the $a_1(1260)$-meson masses are, $m_{D^0}=1.865~{\rm GeV}, m_{D^+}=1.870~{\rm GeV}$ and $m_{a_1} \approx 1.230~{\rm GeV}$, accordingly, which are taken from the Particle Data Group (PDG)~\cite{Zyla:2020zbs}. The $D$-meson and $a_1(1260)$-meson decay constants $f_{D^0}=f_{D^+}=0.210(12)~{\rm GeV}$, $f_{a_1}^\|=0.238(10) ~{\rm GeV}$~\cite{Mutuk:2018lki}. The  $D\to a_1(1260)$ decay processes typical scale in this paper is $\mu_k = (m_D^2 - m_c^2)^{1/2}\approx 1.4~{\rm GeV}$. Furthermore, the non-perturbative vacuum condensates are the significant parameters to the sum rule, and we take~\cite{Colangelo:2000dp,Narison:2014wqa,Zhong:2021epq}
\begin{eqnarray}
\langle \bar qq\rangle |_{2~{\rm GeV}} &=& (-289.14^{+9.34}_{-4.47})^3~{\rm MeV}^3, \nonumber\\
\langle g_s\bar q\sigma TGq\rangle |_{2~{\rm GeV}} &=&(-1.934_{-0.103}^{+0.188})\times  10^{-2}~{\rm GeV}^5, \nonumber\\
\langle g_s\bar qq\rangle^2 |_{2~{\rm GeV}} &=& (2.082_{-0.697}^{+0.734})\times  10^{-3} ~{\rm GeV}^6, \nonumber\\
\langle g_s^2\bar qq\rangle^2 |_\mu &=& (7.420_{-2.483}^{+2.614})\times  10^{-3}~{\rm GeV}^6, \nonumber\\
\langle \alpha_s G^2 \rangle |_\mu &=& 0.038(11)~{\rm GeV}^4, \nonumber\\
\langle g_s^3fG^3\rangle |_\mu &\approx& 0.045 ~{\rm GeV}^6, \nonumber\\
\kappa &=&  0.74(3).
\end{eqnarray}
It should be noted that, the values of the gluon condensates are the most commonly used in QCD sum rules. The value of the double-gluon condensate $\langle \alpha_s G^2 \rangle$ is determined by the sum rule of the charmonium, and the one for triple-gluon condensate $\langle g_s^3fG^3\rangle$ is based on the instanton model\footnote{For more detailed discussion, one can refer to Refs.~\cite{Colangelo:2000dp,Shifman:1978bx,Shifman:1978by}}. The double-quark condensate $\langle \bar qq\rangle$ and the quark-gluon mixed condensate $\langle g_s\bar q\sigma TGq\rangle$ were updated in our previous work~\cite{Zhong:2021epq} based on the GellMann-Oakes-Renner relation and the relationship $\langle g_s\bar q\sigma TGq\rangle = m_0^2 \langle q\bar q\rangle$ with $m_0^2=0.80(2)~{\rm GeV}^2$~\cite{Narison:2014ska}. One can calculate the four-quark condensate $\langle g_s\bar qq\rangle^2$ by using $\rho\alpha_s\langle\bar qq\rangle^2 = (5.8\pm1.8)\times10^{-4}{\rm GeV}^6$ with $\rho \simeq 3-4$~\cite{Narison:2014ska}, and determine the value of $\langle g_s^2\bar qq\rangle^2$ by combining with the new value of $\langle \bar qq\rangle$. All those scale-dependent parameters, such as the quark masses and the vacuum condensates, shall be run from an initial scale $\mu_0$ to a special choice of scale such as $\mu_k$ by applying the renormalization group equations (RGE) given by Refs.~\cite{Yang:1993bp, Hwang:1994vp, Lu:2006fr, Zhang:2021wnv}.

\begin{table}[t]
\begin{center}
\caption{The ratios of the continuum states' and the dimension-six condensates' contributions over the total moments of $a_1(1260)$-meson longitudinal twist-2 LCDA $\xin$ with $n=(2,4,6,8,10)$ within the determined Borel windows. The abbreviations ``Con.'' and ``Six.'' stand for the continuum and dimension-six contributions, respectively.}
\label{tab:m2}
\begin{tabular}{c c c c c c c}
\hline
~~$n$~~ & ~~Con.~~           & ~~~Six.~~~         & ~~~~~$M^2$~~~~~              & ~~~~~$\xin$~~~~~                \\  \hline
2    & $ < 35\%$  & $ < 5\%$ & $[1.782,2.886]$ & $[0.240,0.198]$\\
4    & $ < 35\%$  & $ < 5\%$ & $[2.740,3.385]$ & $[0.101, 0.091]$\\
6    & $ < 40\%$  & $ < 5\%$ & $[3.669,4.567]$ & $[0.057, 0.051]$\\
8    & $ < 40\%$  & $ < 5\%$ & $[4.585,5.269]$ & $[0.037,0.034]$\\
10  & $ < 45\%$  & $ < 5\%$ & $[5.494,6.790]$ & $[0.027,0.023]$   \\  \hline
\end{tabular}
\end{center}
\end{table}

Two important parameters for the BFTSR approach for the moments are continuum threshold $s_0$ and Borel parameter $M^2$, whose range is called as the Borel Window. To fix their values and make the sum rules predictions reliable, the contributions from the continuum states and the contributions from the dimension-six condensates should be small enough. For the purpose, we determine the Borel window by allowing the contribution of continuum states to be less than $45\%$ and the contribution of dimension-six condensates to be less than $5\%$. When determining the threshold parameter $s_{a_1}$, one can normalized the $0_{\rm th}$-order $a_1(1260)$-meson longitudinal DA in the appropriate Borel window. Followed by this approach, we can get $s_{a_1} = 1.4(4)~{\rm GeV}$. We present the determined Borel windows and $\xin$ at the scale $\mu={\sqrt {M^2}}$ in Table~\ref{tab:m2}. Here, we have set the continuum contributions to be no more than $(35\%, 35\%, 40\%, 40\%, 45\%)$ for $n=(2,4,6,8,10)$, respectively, and the dimension-six contributions for all $\xin$ to be less than $5\%$.

\begin{figure}[t]
\includegraphics[width=0.45\textwidth]{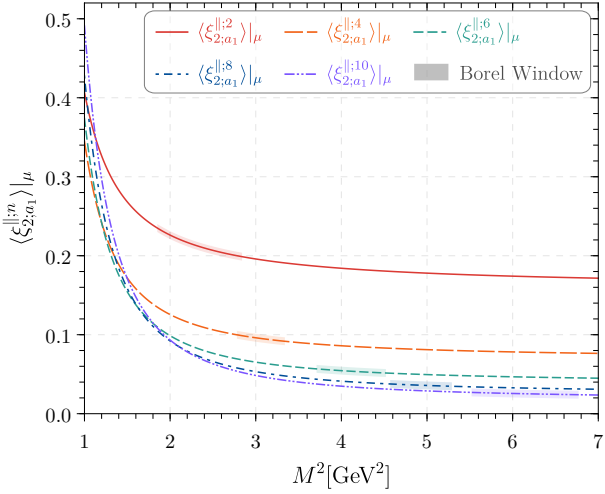}
\caption{Moments $\langle\xi^{\|;n}_{2;a_1}\rangle|_\mu$ up to $n=(2,4,6,8,10)$ order level versus the Borel parameter $M^2$. The shaded bands stand for the corresponding Borel windows.}\label{fig:xi}
\end{figure}

Principally, the moments including all the condensates should be independent to the Borel parameter $M^2$, and for a fixed-order OPE expansion, it may change with different choices of $M^2$ within the allowable Borel window. Such change depends heavily on the convergence of the OPE expansion over $1/M^2$. For the present LCSR up to dimension-6 condensates, e.g the series (\ref{xi2}), as a conservative prediction, we require the variations of $\xin$ within the Borel window to be less than $10\%$. We present the first five moments for $a_1(1260)$-meson twist-2 LCDA versus the Borel parameter $M^2$ in Fig.~\ref{fig:xi}. The shaded bands indicate the corresponding Borel windows, which are in the region of $[1.0,7.0]~{\rm GeV}^2$, respectively.
By taking all uncertainty sources into consideration and applying the RGE of the moments, $\langle\xi_{2;a_1}^{\|;n}\rangle|_\mu$ at the initial scale $\mu_0$ and special scale $\mu_k$ are
\begin{align}
&\langle\xi_{2;a_1}^{\|;2}\rangle|_{\mu_0} = 0.223(29),&&\langle\xi_{2;a_1}^{\|;2}\rangle|_{\mu_k} = 0.219(25),
\nonumber \\
&\langle\xi_{2;a_1}^{\|;4}\rangle|_{\mu_0} = 0.098(8),&&\langle\xi_{2;a_1}^{\|;4}\rangle|_{\mu_k} = 0.097(8),
\nonumber \\
&\langle\xi_{2;a_1}^{\|;6}\rangle|_{\mu_0} = 0.056(6), && \langle\xi_{2;a_1}^{\|;6}\rangle|_{\mu_k} = 0.055(5),
\nonumber \\
& \langle\xi_{2;a_1}^{\|;8}\rangle|_{\mu_0} = 0.039(4),&& \langle\xi_{2;a_1}^{\|;8}\rangle|_{\mu_k} = 0.037(3),
\nonumber \\
& \langle\xi_{2;a_1}^{\|;10}\rangle|_{\mu_0} = 0.028(3), && \langle\xi_{2;a_1}^{\|;10}\rangle|_{\mu_k} = 0.027(3).
\label{daxi}
\end{align}

In order to determine the two LCHO model parameters $\alpha _{2;a_1}^\|$ and $B_{2;a_1}^\|$, one can use the specific fitting by taking the two parameters as the fitting parameters, e.g. $\theta = (\alpha _{2;a_1}^\|, B_{2;a_1}^\|)$. The moments $\langle\xi_{2;a_1}^{\|;n}\rangle|_\mu$ from Eq.~\eqref{Eq:phi} with the definition $\langle\xi_{2;a_1}^{\|;n}\rangle|_\mu = \int_0^1\xi^n\phi_{2;a_1}^\|(x,\mu)$ been regarded as the mean function $\mu(x_i;\theta)(x_i\to n)$, where the moments calculated with BFTSR, i.e. Eq.~\eqref{daxi} are considered as the independent measurements $y_i$ with the known variance $\sigma_i$. To obtain the best values of fitting parameters $\theta$, one can minimize the function
\begin{align}\label{Eq:chi}
  \chi^2(\theta) = \sum\limits_{i=1}^{5}\frac{(y_i - \mu(x_i,\theta))^2}{\sigma_i^2}.
\end{align}
The goodness of fit is judged by the magnitude of the probability
\begin{align}
P_{\chi^2} = \int^\infty_{\chi^2} f(y;n_d) dy.
\label{px2}
\end{align}
Here $f(y;n_d)$ with the number of degrees of freedom $n_d$ is the probability density function of $\chi^2(\theta)$, and
\begin{align}
f(y;n_d) = \frac{1}{\Gamma\left(\dfrac{n_d}{2}\right) 2^{\frac{n_d}{2}}} y^{\frac{n_d}2-1} e^{-\frac y2}.
\end{align}
\begin{figure}[t]
\begin{center}
\includegraphics[width=0.47\textwidth]{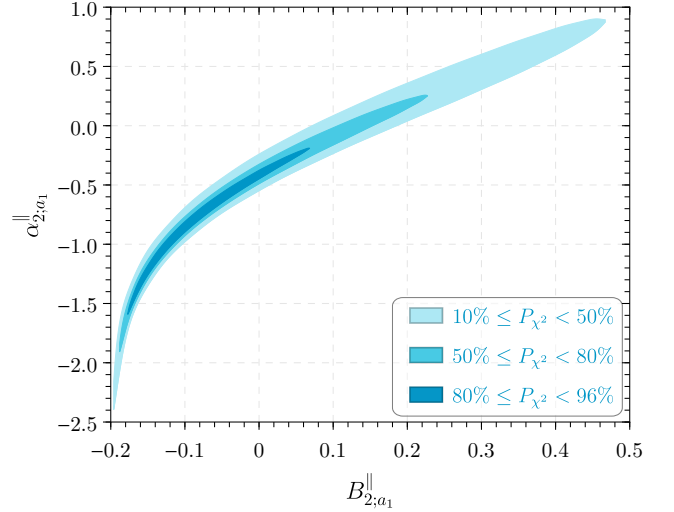}
\end{center}
\caption{The relationship curve between parameters $\alpha _{2;a_1}^\|$, $B_{2;a_1}^\|$ and goodness of fit $P_{\chi_{\min}^2}$.}
\label{fig:fit}
\end{figure}
\begin{table}[b]
\begin{center}
\caption{Fitting parameters $A_{2;a_1}^\|({\rm GeV^{-1}})$, $\beta _{2;a_1}({\rm GeV})$, $\alpha _{2;a_1}^\|$ and $B_{2;a_1}^\|$ with different constituent quark mass $m_q$ under initial scale $\mu_0$. Meanwhile, the goodness of fit $P_{\chi_{\min}^2}$ and the probability $P_{a_1}$ are also given.} \label{tab:phifit}
\begin{tabular}{l c c c c c c c c c}
\hline
$m_q$~~~&~200MeV~&~250MeV~&~300MeV~&~330MeV~&~350MeV~ \\  \hline
$A_{2;a_1}^\|$                  & +6.157        & +5.382     & +4.775     & +4.464        & +4.336       \\
$\beta _{2;a_1}$                & +0.522        & +0.549     & +0.582     & +0.603        & +0.618       \\
$\alpha _{2;a_1}^\|$            & $-0.884$      & $-0.946$   & $-0.996$   & $-1.022$      & $-1.029$  \\
$B_{2;a_1}^\|$                  & $-0.116$      & $-0.115$   & $-0.114$   & $-0.114$      & $-0.113$  \\
$P_{\chi_{\min}^2}$               & +0.977       & +0.954      & +0.923     & +0.903        & +0.889       \\
$P_{a_1}$                         & +0.641       & +0.611      & +0.584     & +0.570        & +0.562      \\ \hline
\end{tabular}
\end{center}
\end{table}
Then, we obtain the fitting parameters at the initial scale $\mu_0$. Due to the quark component of $a_1(1260)$-meson here is $\bar uu$ or $\bar dd$, so the treatment for light-quark mass in this paper is the same with usual constituent quark mass. There are different values for constituent quark mass $m_q$, which is taken to be $250~{\rm MeV}$ in the invariant meson mass scheme~\cite{Jaus:1989au, Chung:1988mu, Choi:1997qh, Schlumpf:1994bc}, $330~{\rm MeV}$ in the spin-averaged meson mass scheme~\cite{Dziembowski:1986dr, Dziembowski:1987zp, Ji:1992yf, Choi:1996mq}. In addition, $m_q=300~{\rm MeV}$ and $m_q=200~{\rm MeV}$~\cite{Zhong:2021epq} of the simplest in Refs.~\cite{Wu:2008yr, Wu:2011gf}. In this work, we present the results for different choices of the constituent quark mass, e.g. $m_q=(200,250,300,330,350)~{\rm MeV}$, respectively. The fitting results are given in Table~\ref{tab:phifit}. As a default value, we shall take $m_q=250~{\rm MeV}$ to do our calculation, whose corresponding goodness of fit is $95.4\%$. This value is also agree with the usual pion and kaon cases~\cite{Zhong:2021epq,Zhong:2022ecl}. One can find that the parameter $A_{2;a_1}^\|$ and $B_{2;a_1}^\|$ gradually decrease with the increment of $m_q$, and the goodness of fit $P_{\chi_{\min}^2}$ is also decreasing with the increment of $m_q$. To show more clearly the relationship between the magnitudes of $\alpha _{2;a_1}^\|$ and $B_{2;a_1}^\|$ and the goodness of fit $P_{\chi_{\min}^2}$, we present the relationship curve between them in Fig.~\ref{fig:fit}. The darker shaded band of Fig.~\ref{fig:fit} represents the higher goodness of fit. When the range of goodness of fit is $80\%  \le P_{\chi^2_{\min}} \le 96\% $, the allowable ranges for the parameters $\alpha_{2;a_1}^\|$ and $B_{2;a_1}^\|$ are quite small.

\begin{figure}[t]
\begin{center}
\includegraphics[width=0.45\textwidth]{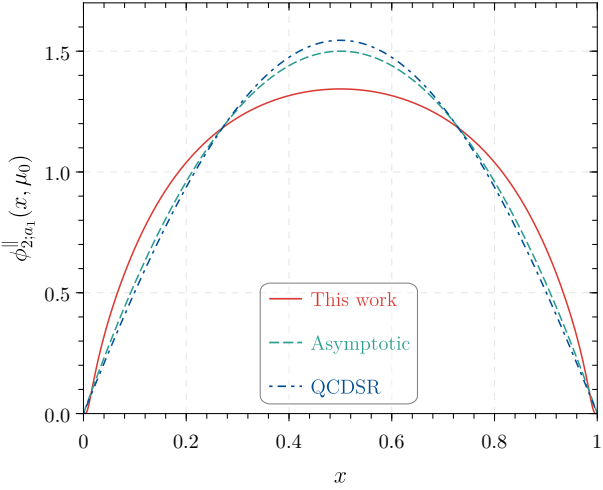}
\end{center}
\caption{The $a_1(1260)$-meson longitudinal twist-2 LCDA $\phi_{2;a_1}^\|(x,\mu_0)$. As a comparison, the QCDSR~\cite{Yang:2008xw, Yang:2007zt} and the asymptotic curves are also given.}
\label{fig:fi}
\end{figure}

\begin{table*}[t]
\begin{center}
\caption{The $D\to a_1(1260)$ TFFs at the large recoil region, i.e. $A(0), V_1(0), V_2(0), V_0(0)$. Predictions under various approaches are also listed.} \label{Tab:TFFs0}
\begin{tabular}{lllll}
\hline
References~~~~~~~~~~~~~~~~~~~~~& $A(0)$ ~~~~~~~~~~~~~~~~~~~~~&$V_1(0)$ ~~~~~~~~~~~~~~~~~~~~~& $V_2(0)$~~~~~~~~~~~~~~~~~~~~~ &  $V_0(0) $   \\   \hline
This~work   & $ 0.130_{-0.013}^{+0.015}$   & $ 1.898_{-0.121}^{+0.128}$    & $ 0.228_{ - 0.021}^{ + 0.020}$    & $ 0.217_{-0.025}^{+0.023}$  \\
LCSR-I~\cite{Momeni:2019uag}    & $0.07\pm0.05$   & $0.37\pm0.01$  & $-0.03\pm0.02$    & $0.15\pm0.05$    \\
LCSR-II~\cite{Huang:2021owr}    & $ 0.34_{-0.04}^{+0.03}$   & $2.63_{-0.21}^{+ 0.20}$  & $0.34_{-0.04}^{+0.03}$    & $0.24_{-0.01}^{+0.00}$    \\
3PSR~\cite{Zuo:2016msr}    & $0.314_{-0.046}^{+0.048}$   & $0.039_{-0.010}^{+0.012}$       & $0.112_{-0.032}^{+0.037}$        & $-0.114_{-0.019}^{+0.018}$  \\
CLFQM~\cite{Cheng:2003sm}    & $0.20$      & $1.54$       & $0.06$        & $0.31$   \\
CLFQM~\cite{Verma:2011yw}   & $0.19_{-0.01}^{+0.00}$   & $1.51_{-0.04}^{+0.00}$  & $0.05_{-0.00}^{+0.01}$    & $0.32_{-0.00}^{+0.00}$    \\
\hline
\end{tabular}
\end{center}
\end{table*}

We present the $a_1(1260)$-meson twist-2 LCDA in Fig.~\ref{fig:fi}. As a comparison, we have also presented the QCDSR prediction~\cite{Yang:2008xw, Yang:2007zt} and the asymmetry behavior $\phi_{2;a_1}^{\|}(x) = 6x\bar x$ in Fig.~\ref{fig:fi}. Our prediction prefers a single-peak behavior, which is as the same as the QCDSR and asymptotic ones. In order to deal with the $a_1(1260)$-meson twist-3 distribution amplitudes, one can decompose the $\phi_{3;a_1}^\bot(x)$ and $\psi_{3;a_1}^\bot(x)$ into several terms according to various source terms~\cite{Ball:1998sk}:
\begin{align}
&\phi_{3;a_1}^\bot(x) = \phi_{3;a_1}^{\bot WW}(x) + \phi_{3;a_1}^{\bot g}(x) + \phi_{3;a_1}^{\bot m}(x) \\
&\psi_{3;a_1}^\bot(x) = \psi_{3;a_1}^{\bot WW}(x) + \psi_{3;a_1}^{\bot g}(x) + \psi_{3;a_1}^{\bot m}(x)
\end{align}
where $\phi_{3;a_1}^{\bot g}(x)$ and $\psi_{3;a_1}^{\bot g}(x)$ are contributions from the three-particle distribution amplitudes, which can be neglected due to their smallest and negligible contributions. $\phi_{3;a_1}^{\bot m}(x)$ and $\psi_{3;a_1}^{\bot m}(x)$ are related to the coefficients $\tilde\delta_\pm$, whose magnitudes tend to be zero in the $q\bar q$ with $q =(u, d)$ meson's components system~\footnote{The detailed analysis can be found in Section 4 in Ref.~\cite{Ball:1998sk}}. The $\phi_{3;a_1}^{\bot WW}(x)$ and $\psi_{3;a_1}^{\bot WW}(x)$ denotes the contribution from the twist-2 longitudinal distribution amplitudes, which is also called Wandzura-Wilczek approximation. Thus, one can get the following relationship
\begin{eqnarray}
\ad \phi_{3;a_1}^\bot(x) = \frac12\bigg[\int_0^x \frac{dv}{\bar v}\phi_{2;a_1}^\|(v) + \int_x^1 \frac{dv}v \phi_{2;a_1}^\|(v)\bigg],
\\
\ad \psi_{3;a_1}^\bot(x) = 2\bar x\int_0^x \frac{dv}{\bar v}\phi_{2;a_1}^\|(v) + 2x\int_x^1 \frac{dv}v \phi_{2;a_1}^\| (v).
\end{eqnarray}
The twist-3,4 LCDAs for $a_1(1260)$-meson is taken from Ref.~\cite{Yang:2007zt}
\begin{eqnarray}
\ad \psi_{4;a_1}^\|(x) = 6x\bar x + (1-3\xi^2)\bigg[\frac17 a_{2;a_1}^{\|;2}  - \frac{20}3 \frac{f_{3;a_1}^A}{f_{a_1}^\| m_{a_1}}\bigg],
\\
\ad \Phi_{3;a_1}^\|(\alpha_i )=5040(\alpha_1-\alpha_2)\alpha_1\alpha_2\alpha_3^2,
\\
\ad \tilde\Phi_{3;a_1}^\| (\alpha_i)=360\alpha_1\alpha_2\alpha_3^2\bigg[1+\frac{1}2(7\alpha_3-3)\omega_{a_1}^V\bigg]
\end{eqnarray}
where $a_{2;a_1}^{\|;2}=0.056\pm 0.072$ is the result obtained by using the BFTSR, the coupling constants $f_{3;a_1}^A=0.0012~{\rm GeV^2}$ and $f_{3;a_1}^V=0.0036~{\rm GeV^2}$~\cite{Yang:2008xw}, the coefficient $\omega_{a_1}^V=-2.9 $~\cite{Yang:2008xw}. $\Phi_{3;a_1}^\|(\alpha_i )$ and $\tilde\Phi_{3;a_1}^\| (\alpha_i)$ represent the twist-3 LCDAs of the three-particle part~\cite{Yang:2007zt, Cheng:2007mx}, accordingly. Here the twist-3 LCDAs $\phi_{3;a_1}^\bot (x)$ and $\psi_{3;a_1}^\bot(x)$ are related to the twist-2 LCDAs by using the Wandzura-Wliczek approximation. In order to evolve the hadronic parameters in $a_1(1260)$-meson twist-2, 3, 4 LCDAs from the initial factorization scale $\mu_0$ to the special scale $\mu_k$, one can use the RGE with the form $ c_i(\mu_k) = L^{\gamma_{c_i}/\beta _0} c_i(\mu_0)$, where $L = \alpha_s(\mu_k)/\alpha_s(\mu_0)$, $\beta_0 = 11 - 2/3 n_f$. The one-loop anomalous dimensions $\gamma_{c_i}$ satisfy the following equation~\cite{Fu:2020uzy}.
\begin{eqnarray}
\gamma_{a_n} = C_F\bigg(1-\frac2{(n+1)(n+2)}-\sum\limits_{m=2}^{n+1} \frac1m \bigg).
\end{eqnarray}

The TFFs of $D\to a_1(1260)$ are key elements for investigating the $D$-meson semileptonic decay. To derive the exact value for the TFFs \eqref{Eq:V1q2}-\eqref{Eq:Aq2}, we need to fix the continuum threshold $s_0$ and Borel parameter $M^2$. Normally, the continuum threshold $s_0$ should be taken near the squared mass of the $D$-meson's first excited state with the same $J^P$ number, e.g. $D_0(2550)^0$. And we take $s_0^A  = 6.5(3)~{\rm GeV}^2$, $s_0^{V_1} = 5.7(3)~{\rm GeV}^2$, $s_0^{V_2} = 6.0(3)~{\rm GeV}^2$ and $s_0^{V_0} = 6.0(3)~{\rm GeV}^2$. One can use four criteria of the LCSR approach listed in Ref.~\cite{Hu:2021zmy} to determine the Borel parameters for the four TFFs. And the determined Borel parameters are ${\cal M}^2_A = 4.0(3)~{\rm GeV}^2$ , ${\cal M}^2_{V_1} = 5.4(3)~{\rm GeV}^2$, ${\cal M}^2_{V_2} = 5.0(3)~{\rm GeV}^2$ and ${\cal M}^2_{V_0} = 5.0(3)~{\rm GeV}^2$, respectively. We present the $D\to a_1(1260)$ TFFs at the large recoil region, i.e. $q^2 \to 0~{\rm GeV}^2$ within errors in Table~\ref{Tab:TFFs0}. To make a comparison, the predictions from various approaches are presented, i.e. the LCSR-I, II~\cite{Huang:2021owr, Momeni:2019uag}, 3PSR~\cite{Zuo:2016msr} and CLFQM~\cite{Cheng:2003sm,Verma:2011yw}, respectively. Our predictions are close to the LCSR-II ones. To have a clear look at the uncertainties caused by different input parameters, we present the TFFs as follows
\begin{eqnarray}
A(0)&=& 0.130+(_{-0.005}^{+0.005})_{s_0}+(_{-0.008}^{+0.009})_{{\cal M}^2}+(_{-0.009}^{+0.010})_{m_c f_D}
\nonumber \\
&=& 0.130_{-0.013}^{+0.015}\\
V_1(0)&=&1.898+(_{-0.065}^{+0.060})_{s_0}+(_{-0.002}^{+0.002})_{{\cal M}^2}+(_{-0.103}^{+0.114})_{m_c f_D}
\nonumber \\
&+& (_{-0.001}^{+0.001})_{a_{2;a_1}^{\|;2}} = 1.898_{-0.121}^{+0.128}\\
V_2(0)&=&0.228+(_{-0.006}^{+0.006})_{s_0}+(_{-0.013}^{+0.015})_{{\cal M}^2}+(_{-0.012}^{+0.014})_{m_c f_D}
\nonumber \\
&+& (_{-0.001}^{+0.001})_{a_{2;a_1}^{\|;2}} = 0.228_{-0.021}^{+0.020}\\
V_0(0)&=&0.217+(_{-0.007}^{+0.006})_{s_0}+(_{-0.020}^{+0.018})_{{\cal M}^2}+(_{-0.012}^{+0.013})_{m_c f_D}
\nonumber \\
&=& 0.217_{-0.025}^{+0.023}
\end{eqnarray}

\begin{table}[b]
\caption{The $D\to a_1(1260)$ TFFs at the large recoil region $q^2=0$, in which the twsit-2,3,4 LCDAs' contributions are presented separately.} \label{Tab:twist}
\begin{tabular}{c c c c c}
\hline
~~~~~~~~~        & ~~~$A(q^2)$~~~     & ~~~$V_1(q^2)$~~~   & ~~~$V_2(q^2)$~~~   & ~~~$V_0(q^2)$~~~  \\   \hline
$\Phi_{2;a_1}^\|(u)$    & $/$           & $/$         & $0.191$       & $-0.240$  \\
$\phi_{3;a_1}^\bot(u)$  & $/$           & $2.061$     & $/$           & $0.550$  \\
$\psi_{3;a_1}^\bot(u)$  & $0.130$       & $/$         & $/$           & $/$       \\
$\Psi_{4;a_1}^\|(u)$    & $/$           & $-0.180$    & $0.019$       & $-0.075$   \\
${\cal H}(\alpha_i)$      & $/$           & $0.017$    & $0.018$       & $-0.018$   \\
Total                   & $0.130$       & $1.898$     & $0.228$       & $0.217$    \\
\hline
\end{tabular}
\end{table}

We present the $D\to a_1(1260)$ TFFs at the large recoil region $q^2=0$ in Table~\ref{Tab:twist}, in which the contributions from the DAs with various twist structures are presented. Ref.~\cite{Lu:2021mri} indicates that the corrections from the two-particle higher-twist contributions are ($27\%$-$36\%$), and our twist-4 contribution falls within this margin of error. As for the twist-4 contribution to the form factor $V_0(0)$, its magnitude is about $34.56\%$ of the total result. For $V_1(0)$ and $V_2(0)$, their twist-4 contributions change to $9.48\%$ and $8.33\%$, respectively. For convenience, we use ${\cal H}(\alpha_i)= \tilde\Phi_{3;a_1}^\|(\alpha_i) -\Phi_{3;a_1}^\|(\alpha_i )$ to represent the net contribution of the three-particle twist-3 LCDAs, whose contribution to the TFFs $V_1(0)$, $V_2(0)$ and $V_0(0)$ are $0.90\%$, $7.90\%$ and $8.30\%$, respectively.

\begin{table}[t]
\caption{The masses of low-lying $D$ resonances, coefficients $\alpha_{1,2}$ and $\Delta$ for the TFFs $A(q^2), V_1(q^2), V_2(q^2), V_0(q^2)$, in which all the input parameters are set to be their central values.} \label{Tab:SSE}
\begin{tabular}{c c c c c}
\hline
~~~~~~~~~        & ~~~$A(q^2)$~~~     & ~~~$V_1(q^2)$~~~   & ~~~$V_2(q^2)$~~~   & ~~~$V_0(q^2)$~~~  \\   \hline
$m_{R,i}$ & 2.007 & 2.420 & 2.420 & 1.865 \\
$\alpha_1$    & $-1.640$   & $-5.178$   & $-0.703$   & $-4.384$  \\
$\alpha_2$    & $31.085$   & $231.525$  & $18.574$    & $38.942$  \\
$\Delta$      & $0.004\%$  & $0.002\%$  & $0.020\%$  & $0.020\%$   \\
\hline
\end{tabular}
\end{table}

Theoretically, the LCSR approach for $D\to a_1(1260)$ TFFs are reliable in low and intermediate $q^2$-regions, which can be extrapolated to all the physically allowable region $m_\ell^2 \leq (m_D - m_{a_1})^2 \approx 0.4~{\rm GeV}^2$. In this paper, we mainly consider the simplified series expansion (SSE), which has the following form
\begin{eqnarray}
F_i(q^2) = \frac1{P_i(q^2)}\sum\limits_k \alpha_k [z(q^2)-z(0)]^k,
\end{eqnarray}
with the function $z(t)$ including $t_\pm$, $t_0$ and $t$, whose definition can be found Ref.~\cite{Hu:2021zmy}. $F_i(q^2)$ are the TFFs $A(q^2)$ and $V_{0,1,2}(q^2)$, respectively. In this approach, the simple pole $P_i(q^2)=(1-q^2 / m^2_{R,i})$ accounting for low-lying resonances, instead of Blaschke factor $B(t)$ is more applicable in many processes. Here, the masses of low-lying $D$ resonances are mainly determined by the $J^P$ states. Followed by the Ref.~\cite{Momeni:2020zrb} and PDG values~\cite{Zyla:2020zbs}, we listed the $m_{R,i}$ in Table~\ref{Tab:SSE}. Meanwhile, the fitting quality should satisfied the relationship $\Delta<1\%$, which is defined as
\begin{align}
\Delta  = \frac{\sum\nolimits_t|F_i(t)-F_i^{\rm fit}(t)|}{ \sum\nolimits_t |F_i(t)|}\times  100
\end{align}
where $t \in [0,1/40, \cdots ,40/40]\times  0.28~{\rm GeV}^2$. The fitting parameters $\alpha_i$ for every TFFs and the quality of fit $\Delta$ are also listed in Table~\ref{Tab:SSE}. From which, the $\Delta$ of $D\to a_1(1260)$ TFFs are less than 0.020$\%$.

\begin{table*}
\begin{center}
\caption{The $D \to a_1(1260)\ell^+\nu_\ell$ branching fractions with error. Other predictions are given as a comparison.} \label{tab:bf}
\begin{tabular}{lllll}
\hline
Method~~~~~~~~~& ${\cal B}(D^0\to a_1^-(1260)e^+\nu_e) $~~~~    & ${\cal B}(D^+\to a_1^0(1260)e^+\nu_e)$~~~~
& ${\cal B}(D^0\to a_1^-(1260)\mu^+\nu_\mu)$~~~~   & ${\cal B}(D^+\to a_1^0(1260)\mu^+\nu_\mu)$
\\   \hline
This work    & $(5.261_{-0.639}^{+0.745})\times 10^{-5}$   & $(6.673_{ - 0.811}^{ + 0.947})\times  10^{-5}$ & $(4.732_{-0.590}^{+0.685})\times  10^{-5}$ & $(6.002_{-0.748}^{+0.796})\times  10^{-5}$    \\
LCSR-I~\cite{Momeni:2019uag}   & $(3.58\pm0.52)\times 10^{-5}$ & $(4.73\pm0.63)\times {10^{ - 5}}$  & - & - \\
LCSR-II~\cite{Huang:2021owr}    & $6.90\times  10^{-5}$  & $9.38\times  10^{-5}$   & $6.27\times  10^{-5}$   & $8.52\times  10^{-5}$   \\
3PSR~\cite{Zuo:2016msr}  & $(1.11_{-0.34}^{+0.41})\times 10^{-5}$   & $(1.47_{-0.44}^{+0.55})\times 10^{-5}$  & - & -\\
CLFQM~\cite{Wang:2015cis}    & $4.1\times 10^{-5}$                         & - &    $3.6\times {10^{-5}}$     & - \\
\hline
\end{tabular}
\end{center}
\end{table*}

\begin{figure*}[t]
\includegraphics[width=0.45\textwidth]{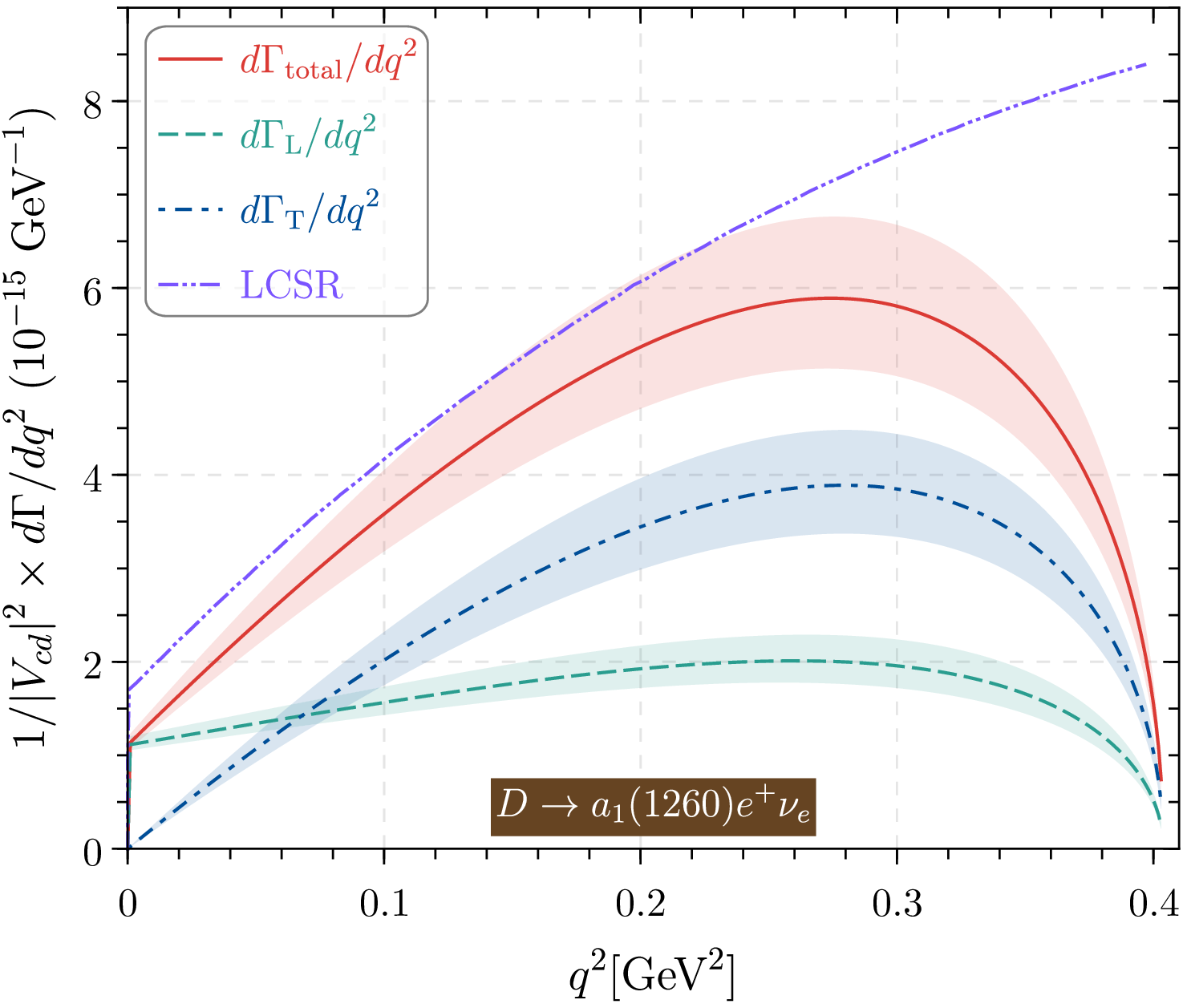}\includegraphics[width=0.45\textwidth]{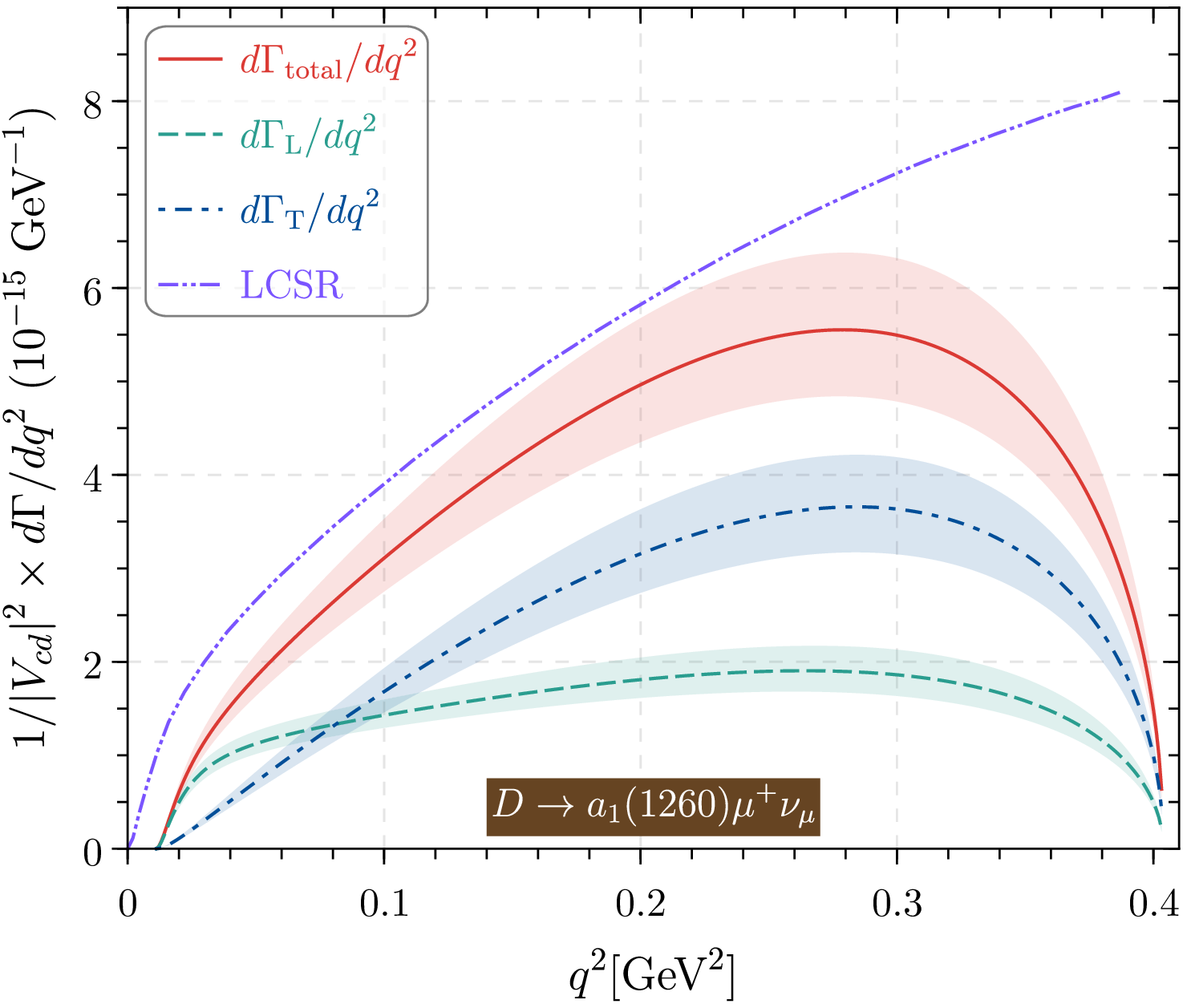}
\caption{The $D \to a_1(1260) \ell^+ \nu_\ell$ decay width with $q^2$ by using the chiral LCSR for the TFFs, where the upper and lower ones are for $\ell= (e,\mu)$, respectively. As a comparison, we also present the other LCSR results from Ref.~\cite{Huang:2021owr}.}
\label{Fig:dG}
\end{figure*}

Furthermore, the $|V_{cd}|$-independence longitudinal and transverse differential decay widths $d\Gamma_{\rm L,T}$ and total width $d\Gamma_{\rm total} = d\Gamma_{\rm L} + d\Gamma_{\rm T}$ of $D \to a_1(1260)\ell^+ \nu_\ell$ can be obtained according to Eqs.~(\ref{Eq:dgamma1}) and (\ref{Eq:dgamma2}), which are shown in Fig.~\ref{Fig:dG}. As a comparison, we also present the LCSR-II predictions~\cite{Huang:2021owr}. Fig.~\ref{Fig:dG} shows that the contributions of the decay widths mainly come from the longitudinal parts in small $q^2$-region, and transverse parts contribute sizably in intermediate and large $q^2$-regions. In different to LCSR-II prediction, our predictions tend to $0$ when $q^2\rightsquigarrow (m_D - m_{a_1})^2 \approx 0.4~{\rm GeV}^2$, which are similar to most of the other semileptonic decay processes such as final state involving $\pi, K, \rho, K^*, D, ...$.

Finally, by using the lifetimes of $D^0, D^+$-mesons ${\tau_{D^0}} = (0.410 \pm 0.015)~\rm ps$ and $\tau_{D^+} = (1.040 \pm 0.007)~\rm ps$ issued by the PDG~\cite{Zyla:2020zbs}, we get the branching fractions for the two different semileptonic decay channels $D^0 \to a_1^-(1260) \ell^+ \nu_\ell$ and $D^+ \to a_1^0(1260) \ell^+ \nu_\ell$, which are listed in Table~\ref{tab:bf}. Meanwhile, other theoretical predictions such as LCSR-I, II~\cite{Momeni:2019uag, Huang:2021owr}, 3PSR~\cite{Zuo:2016msr} and CLFQM~\cite{Wang:2015cis} are also given as a comparison. Our results are in agreement with other predictions which are in order of $10^{-5}$. At present, $D^{0(+)} \to a_1^{-(0)}(1260)\ell^+\nu_\ell$ has not been measured. In year 2020, the BESIII collaboration measured $D^{0(+)} \to b_1^{-(0)}e^+ \nu_e$ decay processes and provided the upper limits for the product branching fractions, which are ${\cal B}(D^0\to b_1^-(1235)e^+\nu_e)\cdot{\cal B}(b_1(1235)^-\to\omega\pi^-) < 1.12\times 10^{-4}$ and ${\cal B}(D^+\to b_1^ 0 (1235)e^+ \nu_e)\cdot{\cal B}(b_1(1235)^0 \to \omega\pi^0 ) < 1.75\times 10^{-4}$~\cite{Ablikim:2020agq}. Here ${\cal B} (b_1(1235)^{0(-)} \to \omega\pi^{0(-)}) = 1$~\cite{Huang:2021owr} and one can get the branching fraction for $D\to b_1(1235)e^+\nu_e $ directly. Furthermore, the branching fractions of $D\to a_1(1260)\ell^+\nu_\ell$ of this work satisfies this upper limit.

\section{Summary}\label{Sec_IV}

In the present paper, we have calculated the $a_1(1260)$-meson moments of LCDA $\xin$ by using the BFTSR approach up to NLO QCD corrections for the perturbative part and up to dimension-six condensates for the non-perturbative part. The moments of LCDA up to $10_{\rm th}$-order have been given in Eq.(\ref{daxi}). Then, by combining the two constraints (\ref{Eq:nc}) and (\ref{Eq:qq}) with the least squares fitting approach for $\xin$, we get the $a_1(1260)$-meson longitudinal LCDA $\phi_{2;a_1}^\|(x,\mu_0)$. Figure~\ref{fig:fi} shows that $\phi_{2;a_1}^\|(x,\mu_0)$ tends to a single-peak behavior. Moreover, by using the derived twist-2 LCDA, we have calculated the $D\to a_1(1260)$ TFFs $A(q^2)$ and $V_{0,1,2}(q^2)$ by using the LCSR approach up to twist-4 accuracy. Furthermore, the $|V_{cd}|$-independence differential decay width of semileptonic decay $D \to a_1(1260)\ell^+\nu_\ell$ with $\ell = (e,\mu)$ have been given in Fig.~\ref{Fig:dG}, and the branching fractions for $D^{0(+)} \to a_1^{-(0)}\ell^+\nu_\ell$ are given in Table~\ref{tab:bf}. The branching fractions are of order $10^{-5}$, which is close to the present experimental upper limit. It is hoped that the decays $D\to a_1(1260)\ell^+\nu_\ell$ can be observed in near future, which inversely could provide a (potential) helpful test for QCD sum rules approach.

\section{Acknowledgments}

We are grateful for the referee's valuable comments and suggestions. This work was supported in part by the National Natural Science Foundation of China under Grant No.11765007, No.11875122, No.11947406, No.12147102 and No.12175025, the Project of Guizhou Provincial Department of Science and Technology under Grant No.KY[2019]1171, and No.ZK[2021]024, the Project of Guizhou Provincial Department of Education under Grant No.KY[2021]030 and No.KY[2021]003, the Fundamental Research Funds for the Central Universities under Grant No.2020CQJQY-Z003, the graduate research and innovation foundation of Chongqing, china (No.ydstd1912), and the Project of Guizhou Minzu University under Grant No. GZMU[2019]YB19.

\end{document}